\documentclass[aps,prd,10pt,twocolumn,superscriptaddress,longbibliography,nofootinbib,floatfix]{revtex4-2}

\usepackage{amsmath,amssymb,bm}
\usepackage{graphicx}
\graphicspath{{figures/}}
\usepackage{xcolor}
\usepackage[hidelinks]{hyperref}

\newcommand{\GeV}{\mathrm{GeV}}
\newcommand{\MeV}{\mathrm{MeV}}
\newcommand{\TeV}{\mathrm{TeV}}

\begin{document}

\title{Unitarity dressing of the dynamical gluon mass scale}

\author{T. V. Iser}
\affiliation{Instituto de F\'isica, Universidade Federal do Rio Grande do Sul, Porto Alegre, RS, Brazil}

\author{V. Li}
\affiliation{Instituto de F\'isica, Universidade Federal do Rio Grande do Sul, Porto Alegre, RS, Brazil}

\author{E. G. S. Luna}
\affiliation{Instituto de F\'isica, Universidade Federal do Rio Grande do Sul, Porto Alegre, RS, Brazil}

\date{\today}

\begin{abstract}
We examine how $s$-channel unitarization changes the dynamical gluon mass scale $m_g$ inferred from high-energy elastic $pp$ scattering.
The Born input is the Reggeized Landshoff--Nachtmann exchange of two
nonperturbative gluons in a color-singlet state.  We fit the ATLAS and
TOTEM forward data separately, using eikonal and $U$-matrix
unitarizations with logarithmic and power-law running masses derived from Schwinger--Dyson analyses.
At the common reference cut
$|t|_{\rm max}=0.12~\GeV^2$, the Born fits give
$m_g=297$--$427~\MeV$; the eikonal and $U$-matrix fits give
$265$--$360~\MeV$ and $243$--$328~\MeV$, respectively.
Unitarization changes the fitted values of the intercept parameter
$\epsilon$, $m_g$, and the exponential slope $a_1$ of the proton form
factor together.
Over $0.10\leq |t|_{\rm max}\leq0.14~\GeV^2$, the $U$-matrix
parameters vary less with the cut, while the eikonal fits to TOTEM
show a pronounced drift.  Replacing the infrared
continuation of the logarithmic effective charge by a Curci--Ferrari
prescription changes $m_g$ by a mean absolute relative shift of $1.77\%$ over
twenty comparisons, with a maximum of $2.17\%$.  The inferred scale
remains in the few-hundred-MeV region, but its fitted value depends on
the nonperturbative input and on the unitarization prescription.
\end{abstract}

\maketitle

\section{Introduction}

An infrared-finite gluon propagator is a characteristic result of
nonperturbative QCD \cite{binosi1,huber1,roberts1}.
Continuum Schwinger--Dyson analyses and large-volume lattice
simulations find that the Landau-gauge gluon two-point function
saturates at small Euclidean momenta
\cite{AguilarBinosiPapavassiliou2008,CucchieriMendes2008,Bogolubsky2009,OliveiraSilva2012}.
This behavior may be described by a momentum-dependent dynamical
gluon mass, generated without an explicit mass term that breaks
gauge symmetry
\cite{Cornwall1982,AguilarNataleRodrigues2003,AguilarBinosiPapavassiliou2011}.
The same framework gives an infrared-finite QCD effective charge.
We ask how the mass scale inferred from hadronic observables
depends on the treatment of $s$-channel unitarity.

Infrared-finite gluon propagators have been used in elastic and
diffractive scattering, forward cross sections, photon-induced
reactions, rapidity-gap survival, and studies of small-$x$ dynamics
\cite{HalzenKreinNatale1993,GayDucatiHalzenNatale1993,LunaEtAl2005,LunaNatale2006,Luna2006,Broilo:2019}.
These applications favor gluon mass scales of a few hundred MeV.
Among the frameworks that allow nonperturbative gluon propagators
to be tested against scattering data, the Landshoff--Nachtmann
(LN) construction is of particular interest at LHC energies,
where Pomeron exchange dominates the forward hadronic amplitude.
The LN model represents the soft Pomeron by a color-singlet pair
of nonperturbative gluons, so that the gluon propagator enters
the elementary scattering amplitude explicitly
\cite{LandshoffNachtmann1987,HalzenKreinNatale1993,GayDucatiHalzenNatale1993,BopsinEtAl2023}.
The explicit LN two-gluon amplitude was used without
$s$-channel unitarization in
Refs.~\cite{HalzenKreinNatale1993,GayDucatiHalzenNatale1993,CudellRoss1991,Canfora:2017,BopsinEtAl2023}.
More recently, a description of the forward ATLAS and TOTEM
differential cross sections at LHC energies was obtained by
Reggeizing the amplitude and choosing a suitable proton
wave-function overlap \cite{BopsinEtAl2023}.
Reggeization supplies the required energy dependence, while
the two-gluon kernel retains the nonperturbative QCD input.

At high energies, $s$-channel unitarity requires multiple exchange
beyond the Born approximation.
These contributions can modify the impact-parameter profile
and the gluon mass scale inferred from elastic scattering.

To the best of our knowledge, the explicit Reggeized LN amplitude
with a dynamical gluon mass has not previously been used as the
Born input for $s$-channel unitarization and then refitted to
high-energy elastic $pp$ data.
We therefore unitarize the amplitude of
Ref.~\cite{BopsinEtAl2023} and examine the resulting changes
in the extracted infrared scale.
We use two standard unitarization schemes, the eikonal and
the $U$-matrix
\cite{BaronePredazzi,Donnachie:2002en,ForshawRoss,Troshin:2007fq}.
With the normalization adopted here, the two schemes share
the same Born term and the same first nonlinear rescattering
correction, but differ in the weights of higher exchanges
and in their large-opacity limits
\cite{Maneyro:2024,LunaRyskin:2024}.

We compare the Born, eikonal, and $U$-matrix descriptions
by fitting $\epsilon$, $m_g$, and $a_1$ to the same data selections.
Here $\epsilon=\alpha_{\mathbb P}(0)-1$ fixes the Pomeron intercept
and governs the energy growth of the Born cross sections.
The mass scale $m_g=m(0)$ is the infrared limit of the running
gluon mass.
The parameter $a_1$ sets the exponential slope of the proton
form factor, $F_1(q^2)=\exp[-a_1q^2]$, where $q^2=-t$.
Varying the upper cut on $|t|$ tests the stability of the
fitted parameters.

The analysis uses forward elastic $pp$ differential cross
sections at $\sqrt{s}=7$, $8$, and $13~\TeV$.
We treat the ATLAS and TOTEM measurements separately because
of the normalization tension between their forward measurements.
The corresponding datasets are denoted by Ensemble A and
Ensemble T, respectively
\cite{ATLAS:2014,ATLAS:2016,ATLAS:2023,TOTEM:2013,TOTEM:2015,TOTEM:2019}.
For the nonperturbative input, we use logarithmic and power-law
running masses derived from Schwinger--Dyson analyses
\cite{AguilarPapavassiliou2008}.

Unitarization changes the fitted values of $\epsilon$, $m_g$,
and $a_1$ together.
The shift in $m_g$ depends on the ensemble.
Over the scanned interval, the $U$-matrix fits show smaller
parameter variation, whereas the eikonal fits to Ensemble T
develop a pronounced cut dependence.
An auxiliary comparison with a Curci--Ferrari prescription
for the effective charge gives much smaller shifts of the
fitted parameters.

The paper is organized as follows.
Section~II gives the Reggeized LN amplitude and the running-mass
parametrizations.
Section~III constructs the eikonal and $U$-matrix profiles.
Section~IV presents the fits, the cut scan, and the comparison
of effective charges.
Section~V contains the conclusions.

\section{Reggeized two-gluon input and dynamical gluon mass scale}
\label{sec:reggeized_input}

The lowest-order QCD exchange with the quantum numbers of the
Pomeron is a pair of gluons in a color-singlet, $C=+1$, state
\cite{Low1975,Nussinov1975,GunionSoper1977,LevinRyskin1981,LandshoffNachtmann1987}.
The massless perturbative gluon propagator has a pole at zero
momentum.
The LN construction replaces it by an infrared-finite
nonperturbative propagator.

Infrared finiteness alone does not supply the observed energy
dependence of elastic scattering.
In the high-energy limit, the unreggeized LN amplitude is
proportional to $s$, so that both the total cross section and
the differential cross section at fixed $t$ are independent
of energy.
Reggeization introduces the factor
$(s/s_0)^{\alpha_{\mathbb P}(t)-1}$ into this amplitude.
The Pomeron intercept governs the growth of the Born cross
sections, while the $t$ dependence of the trajectory determines
the energy evolution of the diffraction cone.
Reggeization is therefore essential to the high-energy
behavior of the LN input.

The calculations of Ref.~\cite{Canfora:2017}, using refined
Gribov--Zwanziger and Cornwall-type propagators, did not
reproduce the ATLAS and TOTEM differential cross sections
at TeV energies, even with an additional phenomenological
energy factor.

To determine how $s$-channel unitarization changes the infrared
scale inferred from elastic scattering, we take the Reggeized
LN amplitude of Ref.~\cite{BopsinEtAl2023} as the Born input.
This amplitude retains the nonperturbative QCD content of
the two-gluon kernel and is given by
\begin{equation}
{\cal M}_B(s,t)
=
i\,\frac{8}{9}\,n_p^2\,
s\left(\frac{s}{s_0}\right)^{\alpha_{\mathbb P}(t)-1}
\left[\widetilde T_1(s,t)-\widetilde T_2(s,t)\right],
\label{eq:born_reggeized}
\end{equation}
where $n_p=3$ is the number of valence quarks in the proton,
$s_0=1~\GeV^2$, and $t=-q^2$.
We use the linear Pomeron trajectory
\begin{equation}
\alpha_{\mathbb P}(t)=1+\epsilon+\alpha'_{\mathbb P}t,
\label{eq:pomeron_trajectory}
\end{equation}
with $\epsilon=\alpha_{\mathbb P}(0)-1$.
We fix $\alpha'_{\mathbb P}=0.25~\GeV^{-2}$, as in standard
soft-Pomeron phenomenology.

In Eq.~\eqref{eq:born_reggeized}, $\widetilde T_1$ contains
the contribution with both gluons coupled to the same quark
in each proton, while $\widetilde T_2$ collects the contributions
involving couplings to different quarks in one or both protons.
They are given by
\begin{widetext}
\begin{eqnarray}
\widetilde T_1
&=&
\int d^2 k\,
\bar\alpha\left(\frac{q}{2}+k\right)
D\left(\frac{q}{2}+k\right)
\bar\alpha\left(\frac{q}{2}-k\right)
D\left(\frac{q}{2}-k\right)
\left[G_p(q,0)\right]^2,
\label{eq:t1}
\end{eqnarray}
\begin{eqnarray}
\widetilde T_2
&=&
\int d^2 k\,
\bar\alpha\left(\frac{q}{2}+k\right)
D\left(\frac{q}{2}+k\right)
\bar\alpha\left(\frac{q}{2}-k\right)
D\left(\frac{q}{2}-k\right)
G_p\left(q,k-\frac{q}{2}\right)
\nonumber\\
&&\times
\left[
2G_p(q,0)-G_p\left(q,k-\frac{q}{2}\right)
\right].
\label{eq:t2}
\end{eqnarray}
\end{widetext}
Here $D$ is the nonperturbative gluon propagator and
$\bar\alpha$ is the corresponding QCD effective charge.
The momentum arguments are transverse and Euclidean.
The proton factor $G_p(q,k)$ is defined by a convolution
of proton wave functions
\cite{BopsinEtAl2023,CudellRoss1991}.
In particular,
\begin{equation}
G_p(q,0)=F_1(q^2),
\label{eq:Gp_F1}
\end{equation}
and, in the forward region, we use
\begin{equation}
F_1(q^2)=\exp[-a_1q^2].
\label{eq:F1_param}
\end{equation}
For the coupling to different quarks, we adopt the
Cudell--Ross prescription \cite{CudellRoss1991},
with a longitudinal quark-momentum fraction of $1/3$,
\begin{equation}
G_p\left(q,k-\frac{q}{2}\right)
=
F_1\left(q^2+9\left|k^2-\frac{q^2}{4}\right|\right).
\label{eq:Gp_approx}
\end{equation}

The combined integrand of $\widetilde T_1-\widetilde T_2$
implements color neutrality: it vanishes when either exchanged
gluon has zero transverse momentum.
The proton wave-function overlap is therefore an essential
part of the LN kernel, together with the two nonperturbative
gluon propagators.

The QCD input appears in Eqs.~\eqref{eq:t1} and \eqref{eq:t2}
through the product $\bar\alpha(Q^2)D(Q^2)$ in the spacelike
region $Q^2>0$.
We use two parametrizations of the running gluon mass derived
from the continuum Schwinger--Dyson analysis of
Ref.~\cite{AguilarPapavassiliou2008}.
That analysis combines the pinch technique with the
background-field method and uses a one-loop dressed
truncation of the gluonic sector.
The dressed three-gluon vertex satisfies the corresponding
Ward identity and contains massless poles, which allow an
infrared-finite propagator while preserving the transversality
of the gluon self-energy.
After separating the effective charge from the massive
propagator, the ultraviolet expansion isolates an integral
equation for $m^2(Q^2)$.
Different leading terms yield the asymptotic behaviors
\begin{equation}
m_{\log}^2(Q^2)
\propto
\left[\ln(Q^2/\Lambda^2)\right]^{-1-\gamma_1},
\end{equation}
and
\begin{equation}
m_{\rm pl}^2(Q^2)
\propto
\frac{1}{Q^2}
\left[\ln(Q^2/\Lambda^2)\right]^{\gamma_2-1}.
\end{equation}
For the vertex Ansatz used in that analysis,
\begin{equation}
\gamma_1 = -\frac{6}{5}(1+c_2-c_1), \qquad
\gamma_2 =\frac{4}{5}+\frac{6}{5}c_1,
\end{equation}
where $c_1$ and $c_2$ are dimensionless coefficients of
the transverse vertex structures.

To extend these asymptotic behaviors to the infrared,
we use the logarithmic form
\begin{equation}
m_{\log}^2(Q^2)
=
m_g^2
\left[
\frac{\ln[(Q^2+\rho_m m_g^2)/\Lambda^2]}
     {\ln[\rho_m m_g^2/\Lambda^2]}
\right]^{-1-\gamma_1},
\label{eq:mlog}
\end{equation}
and the power-law form
\begin{equation}
m_{\rm pl}^2(Q^2)
=
\frac{m_g^4}{Q^2+m_g^2}
\left[
\frac{\ln[(Q^2+\rho_m m_g^2)/\Lambda^2]}
     {\ln[\rho_m m_g^2/\Lambda^2]}
\right]^{\gamma_2-1}.
\label{eq:mpl}
\end{equation}
In Eqs.~\eqref{eq:mlog} and \eqref{eq:mpl}, the shifted
logarithm is finite at the origin, and the normalization
ensures $m_{\log}^2(0)=m_{\rm pl}^2(0)=m_g^2$.
The additional factor $m_g^4/(Q^2+m_g^2)$ in the power-law
form supplies the ultraviolet $1/Q^2$ decrease while
preserving the same infrared limit.
In the original Schwinger--Dyson analysis, $m_g$ and
$\rho_m$ are constrained jointly by the mass condition
\cite{AguilarPapavassiliou2008}.

To limit the number of fit parameters, we fix the dimensionless
mass-shape parameters at
\begin{equation}
\rho_m=4,\qquad
\gamma_1=0.084,\qquad
\gamma_2=2.36,
\label{eq:mass_shape_parameters}
\end{equation}
and take $\Lambda=284~\MeV$ and $n_f=3$.
The value $\rho_m=4$ is the Cornwall choice for the infrared
threshold in the mass function.
The values of $\gamma_1$ and $\gamma_2$ were favored
in an independent analysis of the proton structure function
$F_2(x,Q^2)$, within the ranges allowed by the corresponding
Schwinger--Dyson parametrizations
\cite{LunaNataleDosSantos2011}.
A later analysis, including higher-twist corrections and
different infrared effective charges, retained $\rho_m=4$
and again favored the same values of $\gamma_1$ and $\gamma_2$
\cite{HadjimichefLunaPelaez2020}.
These studies motivate fixing the mass-shape parameters
in the elastic fits, leaving $m_g$ as the only free parameter
within the gluon mass functions.

For every fit reported below, $\rho_m m_g^2>\Lambda^2$, so
the logarithms in the mass functions are well defined.
The fitted values also keep the logarithm in the effective
charge defined below positive throughout the spacelike region.

For either parametrization, $i=\log,{\rm pl}$, the effective
charge is
\begin{equation}
\bar\alpha_i(Q^2)
=
\frac{1}{b_0\ln[(Q^2+4m_i^2(Q^2))/\Lambda^2]},
\label{eq:effective_charge}
\end{equation}
where $b_0=(33-2n_f)/(12\pi)$.
Equivalently,
\begin{equation}
\begin{aligned}
\left[\bar\alpha_i(Q^2)D_i(Q^2)\right]^{-1}
&=b_0\,[Q^2+m_i^2(Q^2)]\\
&\quad\times\ln\left[\frac{Q^2+4m_i^2(Q^2)}{\Lambda^2}\right].
\end{aligned}
\label{eq:alphaD}
\end{equation}
The parameter $m_g$ controls both the infrared strength and
the transverse range of the elementary two-gluon exchange.
The infrared saturation of the gluon propagator is consistent
with lattice results and with subsequent continuum studies
\cite{AguilarBinosiPapavassiliou2008,Ferreira:2025}.
The value of $m_g$ extracted from the scattering fits is
not identified with a lattice screening mass.

The infrared continuation of the QCD coupling is not unique.
An alternative is provided by the massive Landau-gauge
Curci--Ferrari framework.
One- and two-loop calculations in this model describe gluon
and ghost correlation functions obtained from lattice
Yang--Mills simulations
\cite{TissierWschebor2011,GraceyEtAl2019}.
Calculations with dynamical quarks also describe the
corresponding QCD lattice dressing functions in the
two-flavor case \cite{BarriosEtAl2021}.
These comparisons motivate testing the massive-logarithm
charge in Eq.~\eqref{eq:effective_charge} against the
Curci--Ferrari prescription used in
Ref.~\cite{HadjimichefLunaPelaez2020},
\begin{equation}
\bar\alpha_{\rm CF}(Q^2)
=
\frac{\bar\alpha_{\log}(Q^2)}
{1+c_0\ln\left[1+4m_{\log}^2(Q^2)/Q^2\right]},
\label{eq:alphaCF}
\end{equation}
with $c_0=C_A/(24\pi)$ and $C_A=3$.
This prescription incorporates the logarithmic decrease
of the coupling in the infrared-safe renormalization scheme
of the Curci--Ferrari model.
The massive-logarithm charge instead approaches a finite
nonzero value.
In the auxiliary fits, we retain the logarithmic mass and
propagator parametrizations and replace the effective charge.
The three fit parameters are then redetermined.

We use the amplitude normalization
\begin{eqnarray}
\sigma_{\rm tot}(s)
&=&
\frac{1}{s}\,
\mbox{Im}\,{\cal M}_B(s,t=0),
\label{eq:sigtot_born}
\end{eqnarray}
\begin{eqnarray}
\frac{d\sigma}{dt}(s,t)
&=&
\frac{1}{16\pi s^2}\,
\left|{\cal M}_B(s,t)\right|^2 .
\label{eq:dsdt_born}
\end{eqnarray}
We construct the eikonal and $U$-matrix amplitudes in the same
normalization, using this Reggeized exchange as their common
Born input.

\section{Unitary embedding of the Reggeized two-gluon input}
\label{sec:unitary_embedding}

At high energy the unitarity constraint is conveniently written in
impact-parameter space.  Denoting the elastic profile by
${\cal A}(s,b)$, one has
\begin{equation}
2\,\mbox{Im}\,{\cal A}(s,b)
=
|{\cal A}(s,b)|^2+G_{\rm inel}(s,b),
\label{eq:unitarity_b}
\end{equation}
where $G_{\rm inel}(s,b)$ is real and non-negative and represents the
inelastic channels.  If
$\rho(s,b)=\mbox{Re}\,{\cal A}(s,b)/\mbox{Im}\,{\cal A}(s,b)$, the two
solutions of Eq.~\eqref{eq:unitarity_b} may be written as
\begin{equation}
\mbox{Im}\,{\cal A}(s,b)
=
\frac{
1\pm
\sqrt{1-(1+\rho^2)G_{\rm inel}(s,b)}
}
{1+\rho^2},
\label{eq:unitarity_solutions}
\end{equation}
with
\begin{equation}
0\leq G_{\rm inel}(s,b)\leq (1+\rho^2)^{-1}.
\label{eq:ginel_bound}
\end{equation}
For the purely absorptive input used here, the eikonal profile
remains on the branch with the minus sign in
Eq.~\eqref{eq:unitarity_solutions}. The $U$-matrix profile
lies on the same branch below the black-disk limit,
$\mbox{Im}\,{\cal A}=1$, and follows the branch with the
plus sign above it.

We omit the Regge signature phase, so the Born amplitude is purely
imaginary in the present model.  We write its impact-parameter input as
\begin{equation}
\chi(s,b)=i\,\chi_I(s,b),
\label{eq:chi_chiI}
\end{equation}
with
\begin{equation}
\chi_I(s,b)
=
\frac{1}{4\pi s}
\int_0^\infty q\,dq\,
J_0(bq)\,
\mbox{Im}\,{\cal M}_B(s,-q^2).
\label{eq:chiI_def}
\end{equation}
Here $q=\sqrt{-t}$ and $J_0$ is the Bessel function of the first kind.

For the forward cone considered here, a purely absorptive input is a
suitable approximation.  At high energy and small $|t|$, outside the
Coulomb--nuclear interference region, the elastic amplitude is dominated
by its imaginary part \cite{KhozeMartinRyskin2018}.
If a real part is restored and
\begin{equation}
\rho(s,t)=
\frac{\mbox{Re}\,{\cal M}(s,t)}
     {\mbox{Im}\,{\cal M}(s,t)},
\label{eq:rho_t}
\end{equation}
then
\begin{equation}
|{\cal M}(s,t)|^2
=
[\mbox{Im}\,{\cal M}(s,t)]^2
[1+\rho^2(s,t)].
\label{eq:realpart_correction}
\end{equation}
At fixed imaginary amplitude, the relative correction to the
differential cross section is $\rho^2$.  This estimate does not apply in the
diffractive dip, where the imaginary part passes through zero and the
real part is essential for the nonvanishing cross section
\cite{PereiraFerreira2000}. The present fits are restricted to the forward cone, where
$|\rho(s,t)|\ll1$. We therefore retain the purely absorptive
approximation throughout.
The unitarized momentum-space amplitudes are obtained from
\begin{equation}
{\cal M}_X(s,t)
=
4\pi s\int_0^\infty b\,db\,
J_0(bq)\,
{\cal A}_X(s,b),
\qquad X=E,U,
\label{eq:unit_amplitude}
\end{equation}
where $E$ and $U$ denote the eikonal and $U$-matrix prescriptions,
respectively.  The optical theorem gives
\begin{equation}
\sigma_{\rm tot}(s)
=\frac{1}{s}\,\mbox{Im}\,{\cal M}_X(s,0)
=4\pi\int_0^\infty b\,db\,\mbox{Im}\,{\cal A}_X(s,b).
\label{eq:unit_sigtot}
\end{equation}
The corresponding differential cross section is
\begin{equation}
\frac{d\sigma}{dt}(s,t)
=
\frac{1}{16\pi s^2}
\left|{\cal M}_X(s,t)\right|^2 .
\label{eq:unit_dsdt}
\end{equation}

In the eikonal scheme the profile is
\begin{equation}
{\cal A}_E(s,b)
=
i\left[
1-\exp\left\{i\chi(s,b)\right\}
\right].
\label{eq:eikonal_profile}
\end{equation}
With Eq.~\eqref{eq:chi_chiI}, this becomes the usual absorptive form
$i[1-\exp(-\chi_I)]$.  Its expansion is
\begin{eqnarray}
{\cal A}_E(s,b) &=& -i\sum_{n=1}^{\infty} \frac{[i\chi(s,b)]^n}{n!} \nonumber \\
 &=& \chi + \frac{i}{2}\chi^2 - \frac{1}{6}\chi^3 - \frac{i}{24}\chi^4 + \cdots .
\label{eq:eikonal_series_chi}
\end{eqnarray}
The inelastic overlap function in this representation is
\begin{equation}
G_{\rm inel}^{E}(s,b) = 1-\exp[-2\,\mbox{Im}\,\chi(s,b)] .
\label{eq:ginel_eik}
\end{equation}

In the $U$-matrix scheme the profile is
\begin{equation}
{\cal A}_U(s,b) = \frac{\chi(s,b)}{1-i\chi(s,b)/2}.
\label{eq:umatrix_profile}
\end{equation}
For the absorptive input of Eq.~\eqref{eq:chi_chiI}, this becomes
$i\chi_I/(1+\chi_I/2)$.  Its expansion about $\chi=0$ is
\begin{eqnarray}
{\cal A}_U(s,b) &=& -2i\sum_{n=1}^{\infty} \frac{[i\chi(s,b)]^n}{2^n} \nonumber \\
&=& \chi + \frac{i}{2}\chi^2 - \frac{1}{4}\chi^3 - \frac{i}{8}\chi^4 + \cdots .
\label{eq:umatrix_series_chi}
\end{eqnarray}

The series in Eq.~\eqref{eq:umatrix_series_chi} converges for
$|\chi|<2$; the rational expression in Eq.~\eqref{eq:umatrix_profile}
defines the profile beyond this range.  The Born term and the first
nonlinear rescattering correction are the same in
Eqs.~\eqref{eq:eikonal_series_chi} and
\eqref{eq:umatrix_series_chi}.  The prescriptions differ from the third
order onward and in the large-opacity limit.  For an
absorptive input, $\mbox{Im}\,{\cal A}_E\to1$ in the eikonal scheme,
whereas $\mbox{Im}\,{\cal A}_U\to2$ in the $U$-matrix scheme.

The powers of $\chi$ represent successive exchanges of the
Reggeized LN one-Pomeron block.  In this effective description,
the multiple-exchange series is
\begin{equation}
{\mathbb P}+{\mathbb P}{\mathbb P}
+{\mathbb P}{\mathbb P}{\mathbb P}+\cdots .
\label{eq:pomeron_series}
\end{equation}
The eikonal and $U$-matrix prescriptions assign different weights to
these exchanges in impact-parameter space \cite{luna2026a}.
Reggeization of the elementary exchange and its unitarization are
distinct steps.

Since the products in this series are formed in impact-parameter space,
the corresponding momentum-space terms are transverse convolutions,
not ordinary powers of the Born amplitude at fixed $t$.  The
double-exchange term has schematically the form
\begin{equation}
{\cal M}^{(2)}(s,q)
\propto
\frac{i}{s}
\int d^2 k\,
{\cal M}_B(s,-k^2)\,
{\cal M}_B(s,-|\mathbf q-\mathbf k|^2),
\label{eq:double_exchange_convolution}
\end{equation}
with analogous convolutions at higher orders.  The parameters of the
elementary kernel are fitted after this construction of the unitary
amplitude.

\section{Results and physical interpretation}
\label{sec:results}
\label{sec:interpretation}

We fit forward elastic $pp$ differential cross-section data at
$\sqrt{s}=7$, $8$, and $13~\TeV$, using the same data selection
for the Born, eikonal, and $U$-matrix descriptions. The lowest retained values
of $|t|$ at these energies are, respectively, $0.01100$, $0.01050$, and
$0.00754~\GeV^2$ for Ensemble A, and $0.008180$,
$0.008628$, and $0.006498~\GeV^2$ for Ensemble T.
These lower limits lie outside the Coulomb--nuclear
interference region and are kept fixed. To study the stability
of the unitarized fits, we vary only the upper limit,
\begin{equation}
0.10\leq |t|_{\rm max}\leq0.14~\GeV^2 .
\label{eq:t_scan}
\end{equation}
We take $|t|_{\rm max}=0.12~\GeV^2$ as the common reference cut
for the three descriptions. At this cut, Ensembles A and T
contain $83$ and $218$ data points, respectively, giving
$\nu=80$ and $215$ for the three-parameter fits.
The reference cut precedes the pronounced
drift of the TOTEM eikonal parameters at the two largest cuts.
The full scan is retained to assess parameter stability.

Ensemble A contains the ATLAS differential cross sections and
Ensemble T the TOTEM measurements. For each ensemble, amplitude
prescription, and mass parametrization, we fit $\epsilon$, $m_g$,
and $a_1$ independently. Minimization is performed with MINUIT
\cite{JamesRoos1975}.

The statistical and systematic uncertainties of each data point
are added in quadrature, and correlations between data points
are neglected. Parameter uncertainties are obtained from the
covariance matrix calculated from the second derivatives of
$\chi^2$ at its minimum. Within the local quadratic
approximation, the errors quoted in the tables and shown in
the figures are the coordinate projections of the nominal
joint $90\%$ confidence region for the three fitted parameters,
\begin{equation}
\chi^2-\chi^2_{\rm min}\leq6.251.
\label{eq:confidence_region}
\end{equation}
Fit quality is reported as $\chi^2_{\rm min}/\nu$, where $\nu$
is the number of degrees of freedom.

\begin{figure*}[!t]
\centering
\includegraphics[width=\textwidth]{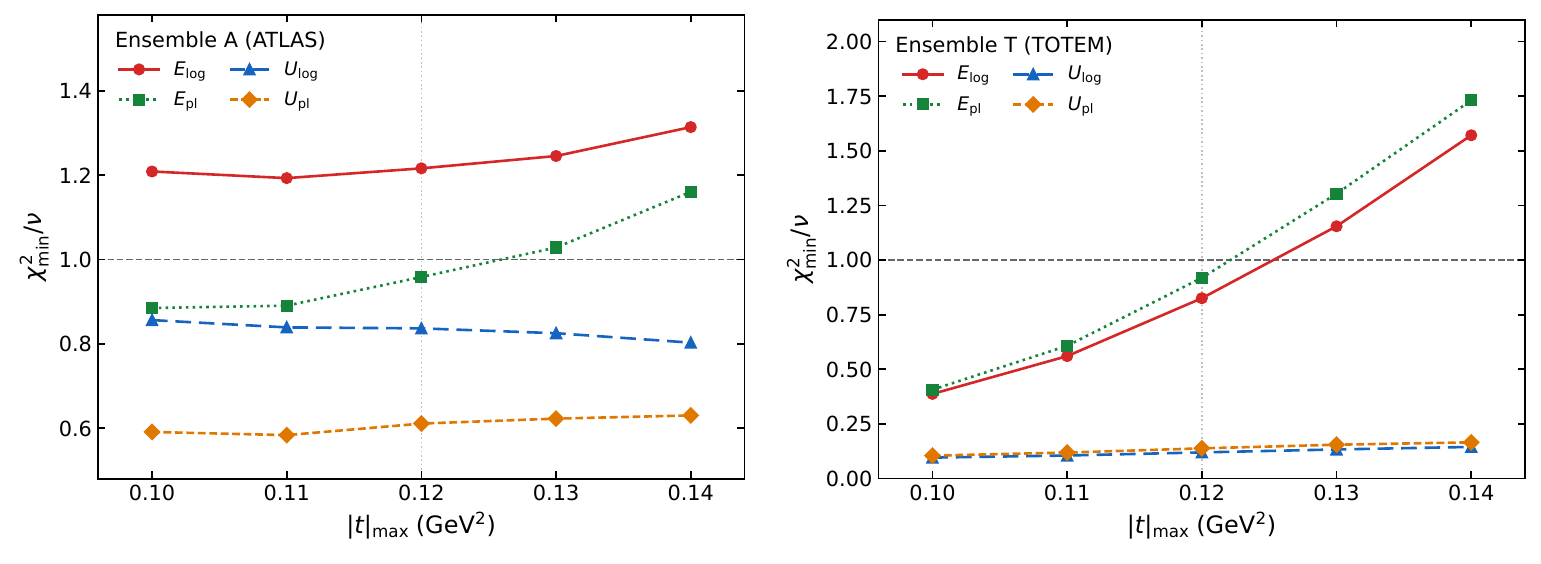}
\caption{Dependence of $\chi^2_{\rm min}/\nu$ on the upper
momentum-transfer limit $|t|_{\rm max}$ for Ensemble A
(ATLAS, left) and Ensemble T (TOTEM, right). The labels $E$
and $U$ denote the eikonal and $U$-matrix schemes, while
$\log$ and $\mathrm{pl}$ denote the logarithmic and power-law
running masses. The horizontal line at $\chi^2/\nu=1$ is
shown for reference.}
\label{fig:cut_scan}
\end{figure*}

Figure~\ref{fig:cut_scan} shows the dependence of the fit
quality on the upper cut. For Ensemble A, all four fits vary
moderately over the scanned interval. For Ensemble T, the
eikonal values rise from $\chi^2/\nu\simeq0.39$--$0.41$ at
$|t|_{\rm max}=0.10~\GeV^2$ to $1.57$--$1.73$ at
$0.14~\GeV^2$, whereas the $U$-matrix values remain between
about $0.10$ and $0.17$. Both schemes use the same
one-parameter proton form factor; the different cut dependence
is obtained after refitting the input parameters under the
two nonlinear maps.

\begin{figure*}[!t]
\centering
\includegraphics[width=\textwidth]{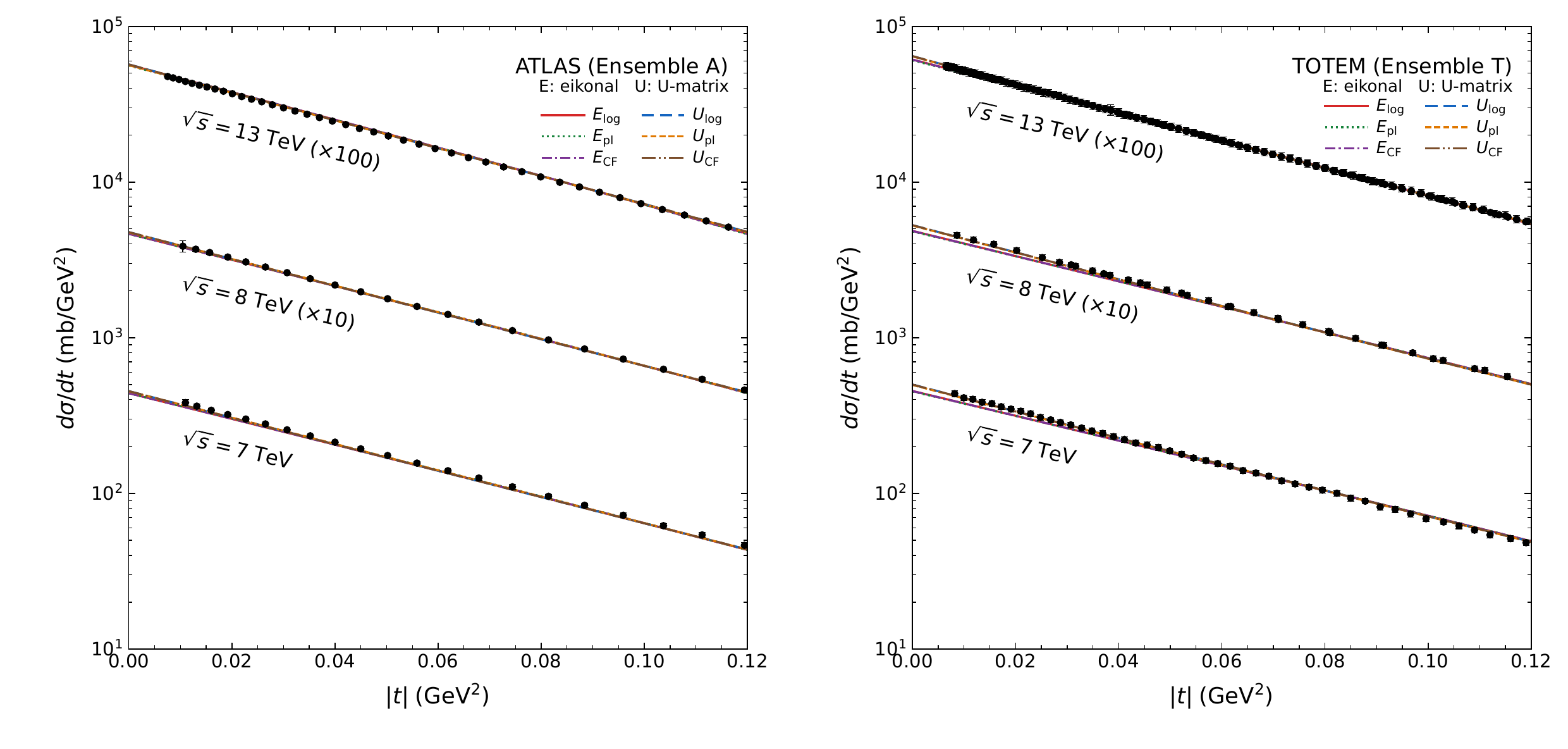}
\caption{Forward elastic $pp$ differential cross sections at
$\sqrt{s}=7$, $8$, and $13~\TeV$ for Ensemble A (ATLAS, left)
and Ensemble T (TOTEM, right). Points show the measurements
with their experimental uncertainties. For each energy, the
curves show the six unitarized model variants evaluated with
the reference-cut parameters of
Tables~\ref{tab:log_results}--\ref{tab:cf_results}.
The data and curves at $8$ and $13~\TeV$ are multiplied by
$10$ and $100$, respectively, for display. The labels $E$
and $U$ denote the eikonal and $U$-matrix schemes;
$\log$ and $\mathrm{pl}$ denote the logarithmic and power-law
running masses. The label $\mathrm{CF}$ denotes the
logarithmic mass combined with the Curci--Ferrari effective
charge of Eq.~\eqref{eq:alphaCF}.}
\label{fig:dsdt}
\end{figure*}

The differential cross sections at the reference cut are shown
in Fig.~\ref{fig:dsdt}. Within each unitarization scheme,
the logarithmic, power-law, and Curci--Ferrari variants give
similar forward curves. The cut dependence of the fitted
parameters is examined below.

\begin{figure*}[t]
\centering
\includegraphics[width=\textwidth]{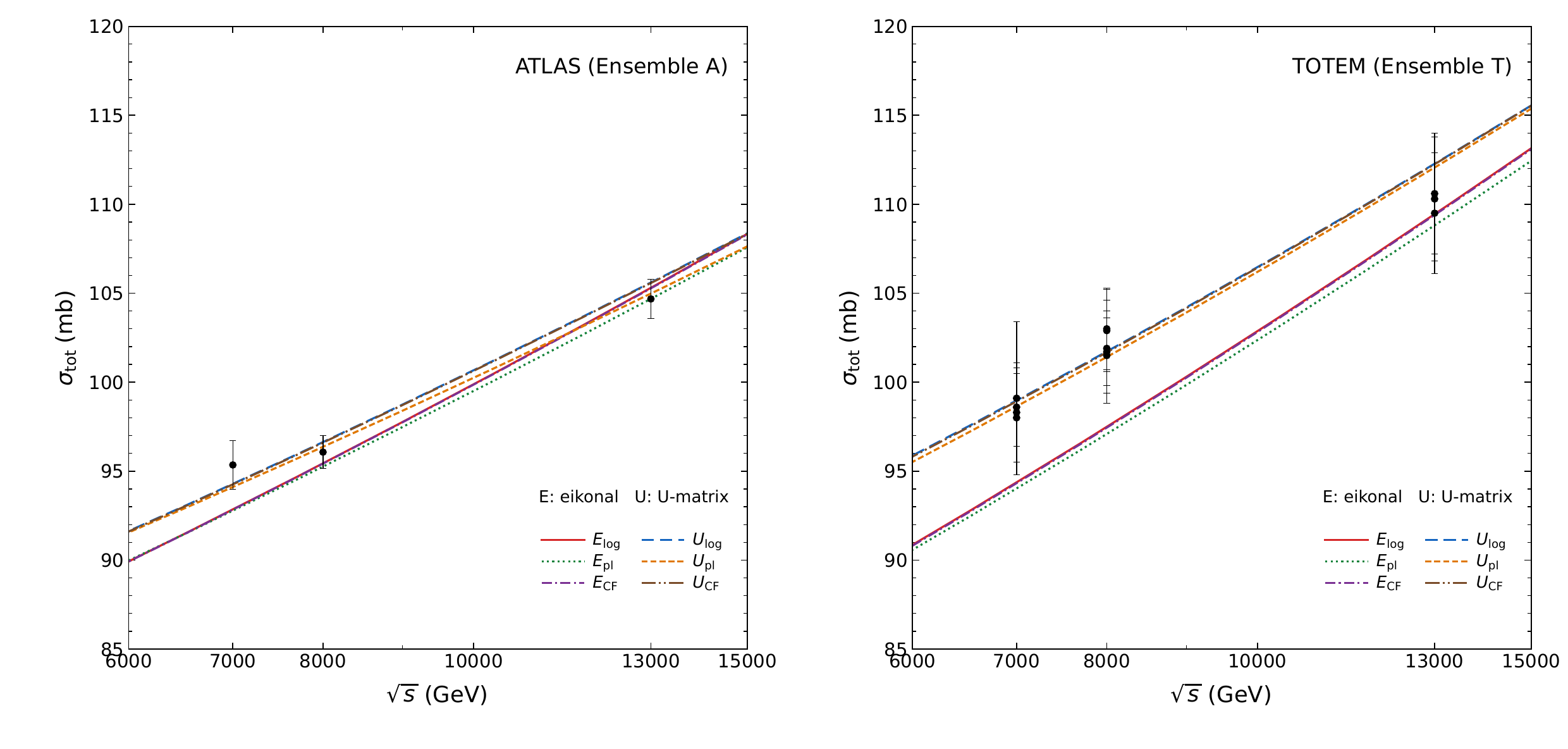}
\caption{Total $pp$ cross sections predicted from the
reference-cut differential cross-section fits for Ensemble A
(ATLAS, left) and Ensemble T (TOTEM, right). Points show the
corresponding total-cross-section measurements with their
experimental uncertainties; these data were not included
in the fits. Model labels and line styles are the same as in
Fig.~\ref{fig:dsdt}.}
\label{fig:sigtot}
\end{figure*}

The same parameter sets determine the total cross section
through Eq.~\eqref{eq:unit_sigtot}. Figure~\ref{fig:sigtot}
compares these predictions with the ATLAS and TOTEM
measurements, which were not included in the fit. No additional
normalization or refitting is applied. The separation between
the eikonal and $U$-matrix curves is more pronounced for
Ensemble T than for Ensemble A, while the Curci--Ferrari
curves remain close to the corresponding logarithmic-charge
curves. The total cross sections thus provide a separate
comparison of the energy dependence selected by the forward
elastic fits.

Tables~\ref{tab:log_results} and \ref{tab:pl_results} collect
the Born, eikonal, and $U$-matrix results at the common
reference cut.

\begin{table*}[!t]
\centering
\caption{Fit parameters obtained with the logarithmic running
mass $m_{\log}(Q^2)$ at $|t|_{\rm max}=0.12~\GeV^2$.
Ensemble A denotes the ATLAS data set and Ensemble T the
TOTEM data set. Parameter uncertainties follow the joint
confidence convention of Eq.~\eqref{eq:confidence_region}.}
\begin{ruledtabular}
\begin{tabular}{lcccccc}
 & \multicolumn{2}{c}{Born}
 & \multicolumn{2}{c}{Eikonal}
 & \multicolumn{2}{c}{$U$-matrix}\\
\cline{2-3}\cline{4-5}\cline{6-7}
 & A & T & A & T & A & T \\
\hline
$m_g$ $(\MeV)$
& $297\pm14$ & $363\pm14$
& $297\pm23$ & $265\pm21$
& $274\pm19$ & $243\pm19$ \\
$\epsilon$
& $0.0623\pm0.0060$ & $0.0928\pm0.0057$
& $0.1374\pm0.0030$ & $0.1704\pm0.0085$
& $0.1119\pm0.0025$ & $0.1272\pm0.0062$ \\
$a_1$ $(\GeV^{-2})$
& $1.34\pm0.11$ & $1.757\pm0.065$
& $0.401\pm0.077$ & $0.199\pm0.049$
& $0.480\pm0.086$ & $0.302\pm0.076$ \\
\hline
$\nu$
& 80 & 215 & 80 & 215 & 80 & 215 \\
$\chi^2/\nu$
& 0.71 & 0.24 & 1.22 & 0.83 & 0.84 & 0.12 \\
\end{tabular}
\end{ruledtabular}
\label{tab:log_results}
\end{table*}

\begin{table*}[!t]
\centering
\caption{Fit parameters obtained with the power-law running
mass $m_{\rm pl}(Q^2)$ at $|t|_{\rm max}=0.12~\GeV^2$.
The ensemble labels and uncertainty convention are those
of Table~\ref{tab:log_results}.}
\begin{ruledtabular}
\begin{tabular}{lcccccc}
 & \multicolumn{2}{c}{Born}
 & \multicolumn{2}{c}{Eikonal}
 & \multicolumn{2}{c}{$U$-matrix}\\
\cline{2-3}\cline{4-5}\cline{6-7}
 & A & T & A & T & A & T \\
\hline
$m_g$ $(\MeV)$
& $360\pm19$ & $427\pm15$
& $360\pm30$ & $309\pm29$
& $328\pm28$ & $267\pm20$ \\
$\epsilon$
& $0.0628\pm0.0064$ & $0.0887\pm0.0050$
& $0.1311\pm0.0032$ & $0.1676\pm0.0080$
& $0.1077\pm0.0030$ & $0.1288\pm0.0062$ \\
$a_1$ $(\GeV^{-2})$
& $1.46\pm0.13$ & $1.813\pm0.067$
& $0.438\pm0.090$ & $0.198\pm0.031$
& $0.516\pm0.072$ & $0.273\pm0.050$ \\
\hline
$\nu$
& 80 & 215 & 80 & 215 & 80 & 215 \\
$\chi^2/\nu$
& 0.50 & 0.15 & 0.96 & 0.92 & 0.61 & 0.14 \\
\end{tabular}
\end{ruledtabular}
\label{tab:pl_results}
\end{table*}

\begin{figure*}[!t]
\centering
\includegraphics[width=\textwidth]{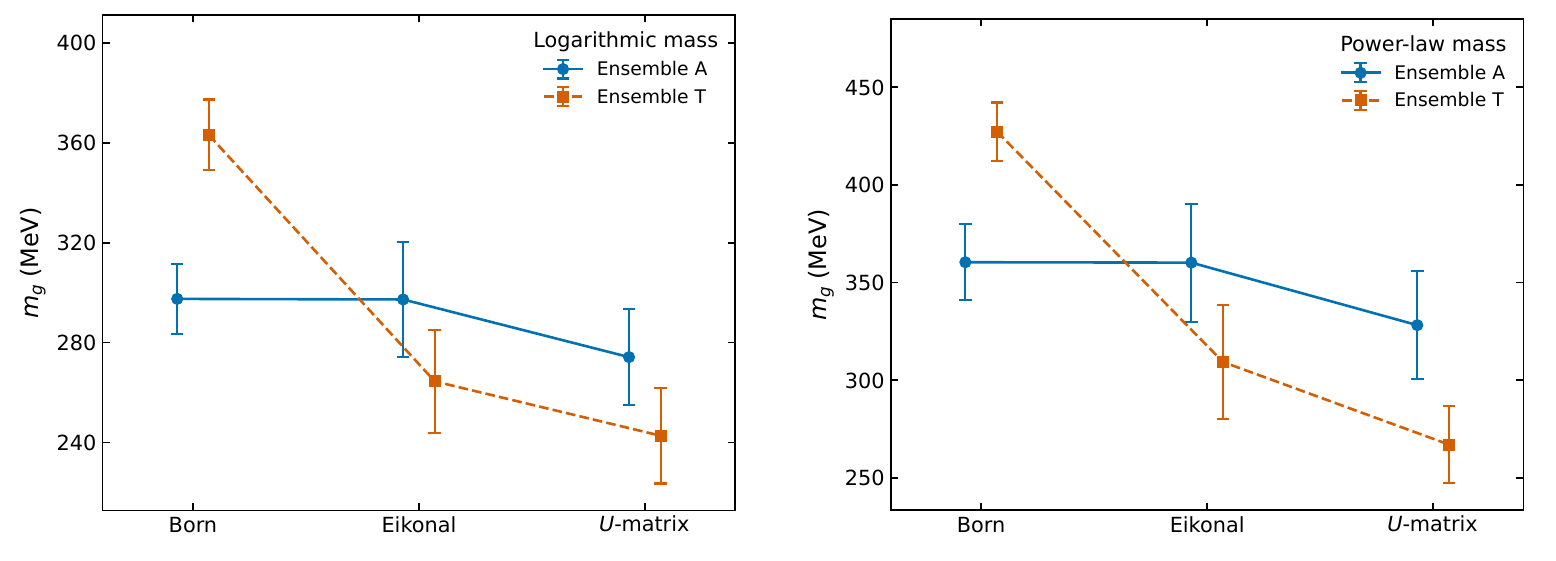}
\caption{Fitted gluon mass scale at
$|t|_{\rm max}=0.12~\GeV^2$ for the logarithmic (left)
and power-law (right) running masses. Born, eikonal, and
$U$-matrix results are shown for Ensembles A and T.
The error bars follow the joint confidence convention
of Eq.~\eqref{eq:confidence_region}.}
\label{fig:mg_summary}
\end{figure*}

The central values of the mass scale lie in the ranges
\begin{equation}
297~\MeV \lesssim m_g^B \lesssim 427~\MeV,
\end{equation}
\begin{equation}
265~\MeV \lesssim m_g^E \lesssim 360~\MeV,
\end{equation}
and
\begin{equation}
243~\MeV \lesssim m_g^U \lesssim 328~\MeV.
\end{equation}
All three descriptions therefore retain the dynamical gluon
mass in the few-hundred-MeV domain.
Figure~\ref{fig:mg_summary} summarizes these mass scales.

To quantify the change of the infrared scale, we define
\begin{equation}
R_E=\frac{m_g^E}{m_g^B},
\qquad
R_U=\frac{m_g^U}{m_g^B}.
\label{eq:mass_ratios}
\end{equation}
For the logarithmic mass,
\begin{eqnarray}
R_E^A&=&0.999, \qquad R_U^A=0.922,\\
R_E^T&=&0.728, \qquad R_U^T=0.669,
\label{eq:log_ratios}
\end{eqnarray}
and for the power-law mass,
\begin{eqnarray}
R_E^A&=&0.999, \qquad R_U^A=0.911,\\
R_E^T&=&0.724, \qquad R_U^T=0.625.
\label{eq:pl_ratios}
\end{eqnarray}
The two mass parametrizations give nearly the same relative
response to eikonal unitarization in each ensemble. For
Ensemble A, the eikonal mass is essentially unchanged,
while the $U$-matrix lowers it by about $8$--$9\%$.
For Ensemble T, the reduction is about $27\%$ in the
eikonal scheme and $33$--$38\%$ in the $U$-matrix scheme.

Unitarization also changes $\epsilon$ and $a_1$. For Ensemble A,
the logarithmic fit changes from
$(\epsilon,m_g,a_1)=(0.0623,297~\MeV,1.34~\GeV^{-2})$
at Born level to $(0.1374,297~\MeV,0.401~\GeV^{-2})$
in the eikonal scheme and
$(0.1119,274~\MeV,0.480~\GeV^{-2})$ in the $U$-matrix
scheme. The power-law mass shows the same pattern.
The nearly unchanged eikonal mass thus accompanies
substantial changes of the energy dependence and proton
overlap. For Ensemble T, the reductions of $m_g$ accompany
an increase of $\epsilon$ and an even larger relative
decrease of $a_1$. We use ``unitarity dressing'' to denote
this joint readjustment of the three parameters when the
Born kernel is refitted within a unitary amplitude.

At the reference cut, the Born values of $\chi^2/\nu$ lie
between $0.15$ and $0.71$, the $U$-matrix values between
$0.12$ and $0.84$, and the eikonal values between $0.83$
and $1.22$. These absolute values depend on the adopted
error prescription and are considered together with the
cut dependence in Fig.~\ref{fig:cut_scan}.

The central values in all four eikonal/$U$-matrix comparisons
satisfy
\begin{equation}
\epsilon_U<\epsilon_E,
\qquad
m_{g,U}<m_{g,E},
\qquad
a_{1,U}>a_{1,E}.
\label{eq:ordering}
\end{equation}
This ordering follows from refitting the common elementary
input under the two nonlinear maps. The smaller $\epsilon_U$
implies a weaker energy dependence of the Born input in the
$U$-matrix fit, together with different transverse parameters.
The larger limiting imaginary profile allowed by this
scheme does not, by itself, establish that the limit is
approached in the fitted energy range.

\begin{table*}[!t]
\centering
\caption{Fit parameters obtained with the Curci--Ferrari
effective charge of Eq.~\eqref{eq:alphaCF} and the
logarithmic mass and propagator parametrizations at
$|t|_{\rm max}=0.12~\GeV^2$. Parameter uncertainties follow
Eq.~\eqref{eq:confidence_region}. The last row gives the
relative shift of $m_g$ with respect to the corresponding
logarithmic-charge fit, as defined in Eq.~\eqref{eq:cf_shift}.}
\begin{ruledtabular}
\begin{tabular}{lcccc}
 & $A,E$ & $A,U$ & $T,E$ & $T,U$ \\
\hline
$\epsilon$
& $0.1373\pm0.0030$ & $0.1120\pm0.0025$
& $0.1704\pm0.0085$ & $0.1276\pm0.0057$ \\
$m_g$ $(\MeV)$
& $291\pm22$ & $269\pm18$
& $260\pm28$ & $239\pm19$ \\
$a_1$ $(\GeV^{-2})$
& $0.402\pm0.077$ & $0.482\pm0.086$
& $0.198\pm0.049$ & $0.300\pm0.068$ \\
$\chi^2/\nu$
& $1.21$ & $0.84$ & $0.84$ & $0.12$ \\
$\delta_{m_g}^{\rm CF}$
& $-2.05\%$ & $-1.76\%$ & $-1.79\%$ & $-1.67\%$ \\
\end{tabular}
\end{ruledtabular}
\label{tab:cf_results}
\end{table*}

\begin{figure*}[!t]
\centering
\includegraphics[width=\textwidth]{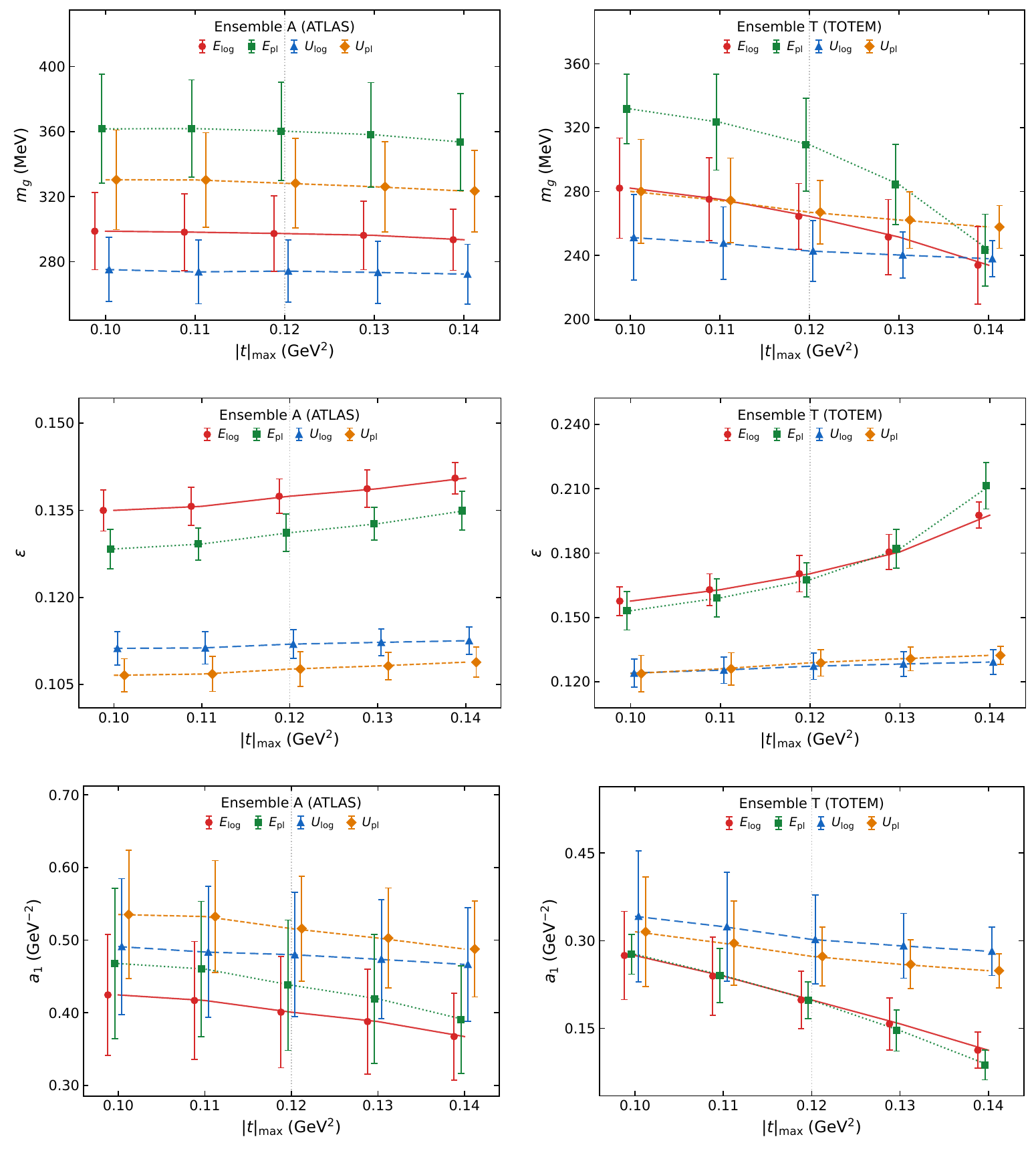}
\caption{Dependence of $m_g$ (top), $\epsilon$ (middle), and
$a_1$ (bottom) on $|t|_{\rm max}$ for Ensemble A (left column)
and Ensemble T (right column). Model labels are the same as
in Fig.~\ref{fig:cut_scan}. Error bars follow
Eq.~\eqref{eq:confidence_region}, and lines join the central
values at the fitted cuts. Fits at different cuts share
data points and are therefore correlated.}
\label{fig:param_scan}
\end{figure*}

Figure~\ref{fig:param_scan} shows the parameter dependence
on the upper cut. In Ensemble A, varying the cut from $0.10$
to $0.14~\GeV^2$ changes $m_g$ by only a few percent in all
four fits. In Ensemble T, the $U$-matrix masses decrease
moderately, from $251$ to $238~\MeV$ for the logarithmic
mass and from $280$ to $258~\MeV$ for the power-law mass.
The eikonal variation is stronger: the logarithmic mass
decreases from $282$ to $234~\MeV$, and the power-law mass
from $332$ to $243~\MeV$. The accompanying increase of
$\epsilon$ and decrease of $a_1$ are even more pronounced.
This parameter drift accompanies the deterioration of
$\chi^2/\nu$ in Fig.~\ref{fig:cut_scan}.

\subsection{Infrared-coupling sensitivity}
\label{sec:coupling_sensitivity}

The logarithmic and power-law mass fits use the same
massive-logarithm prescription for the effective charge.
We test its infrared continuation by repeating the
logarithmic fits with the Curci--Ferrari charge of
Eq.~\eqref{eq:alphaCF}. The logarithmic mass and propagator
parametrizations are retained, and all three parameters
are refitted.

For a quantity $X$, the relative shift in percent is
\begin{equation}
\delta_X^{\rm CF}
=
100\,\frac{X_{\rm CF}-X_{\log}}{X_{\log}}.
\label{eq:cf_shift}
\end{equation}
Table~\ref{tab:cf_results} gives the Curci--Ferrari results
at the reference cut.

The gluon mass is reduced by $2.05\%$, $1.76\%$, $1.79\%$,
and $1.67\%$ in the four ATLAS/TOTEM and eikonal/$U$-matrix
comparisons, respectively. The relative shifts of $\epsilon$,
$a_1$, and $\chi^2/\nu$ have magnitudes below $0.26\%$,
$0.78\%$, and $2.5\%$, respectively.

The same pattern holds over the full scan. For the twenty
Curci--Ferrari/logarithmic pairs in
$0.10\leq|t|_{\rm max}\leq0.14~\GeV^2$, the mean absolute
relative shift of $m_g$ is $1.77\%$, and the largest is
$2.17\%$. The corresponding means are $0.13\%$ for
$\epsilon$, $0.54\%$ for $a_1$, and $1.08\%$ for
$\chi^2/\nu$; the largest absolute relative shift of $a_1$
is $1.77\%$.

The changes between these two infrared coupling prescriptions
are smaller than those produced by the mass parametrization
and, in particular, by unitarization of the LN kernel.
We treat these shifts as an estimate of model dependence
rather than as a contribution to the statistical parameter
uncertainties.

\clearpage
\section{Conclusions}
\label{sec:conclusions}

We have compared the Reggeized LN Born amplitude with its
eikonal and $U$-matrix unitarizations using forward elastic
$pp$ data at $\sqrt{s}=7$, $8$, and $13~\TeV$.
The logarithmic and power-law running masses were tested
with the same data selections in all three descriptions,
treating ATLAS and TOTEM separately.

At $|t|_{\rm max}=0.12~\GeV^2$, the logarithmic and power-law
fits with the massive-logarithm charge give mass scales
between $243$ and $427~\MeV$. Their response to unitarization
depends on the ensemble. For ATLAS, the eikonal mass is
nearly unchanged relative to the Born fit, while the
$U$-matrix lowers it by about $8$--$9\%$. For TOTEM,
the reductions are about $27\%$ and $33$--$38\%$,
respectively. In both ensembles, $\epsilon$ increases
and $a_1$ decreases relative to the Born fits.

The cut scan distinguishes the two unitarization schemes
most clearly for TOTEM. Over
$0.10\leq|t|_{\rm max}\leq0.14~\GeV^2$,
the $U$-matrix parameters vary less, whereas the TOTEM
eikonal fits develop pronounced parameter drift and
a rapid increase of $\chi^2/\nu$.

The total cross sections predicted from the differential
fits provide a comparison with measurements omitted from
the minimization. The eikonal and $U$-matrix predictions
lie closer to one another for ATLAS than for TOTEM.
Replacing the massive-logarithm charge with the
Curci--Ferrari prescription gives a mean absolute relative
shift of $1.77\%$ in $m_g$ over the full scan, with a
maximum of $2.17\%$, and preserves the main pattern
of cut dependence.

Within the one-channel descriptions and fixed mass-shape
parameters considered here, the infrared scale inferred
from elastic scattering depends on the treatment of
$s$-channel unitarity. Its extraction requires a joint
determination of the mass scale, the energy dependence
of the Born input, and the proton overlap.

\begin{acknowledgments}

The authors thank M.~N.~Ferreira, A.~A.~Natale and M.~G.~Ryskin for useful discussions.
This research was partially supported by the Coordena\c{c}\~ao de
Aperfei\c{c}oamento de Pessoal de N\'{\i}vel Superior (CAPES) and by the
Conselho Nacional de Desenvolvimento Cient\'{\i}fico e Tecnol\'ogico
(CNPq).

\end{acknowledgments}

\end{document}